\documentclass[reprint,amsmath,amssymb,aps,prl,floatfix,groupedaddress,superscriptaddress]{revtex4-2}
\usepackage{graphicx}
\usepackage{comment}
\usepackage[caption=false]{subfig}
\usepackage{dcolumn}
\usepackage{bm}
\usepackage[mathlines]{lineno}
\usepackage{mathtools}
\usepackage{physics}
\usepackage{times,newtx}
\usepackage{hyperref}
\usepackage{color}
\usepackage{overpic}

\begin{document}


\title{Quantum Limits of Passive Optical Surface Metrology and Defect Detection}

\author{Jernej Frank}
\email{jernej.frank@physics.ox.ac.uk}
\affiliation{%
 Department of Physics, University of Oxford, Oxford, OX1 3PU, UK
}

\author{George Brumpton}
\email{pmygb8@nottingham.ac.uk}
\affiliation{%
 {Manufacturing~Metrology~Team,~Faculty~of~Engineering,~University~of~Nottingham,~Nottingham~NG7~2RD,~UK}
}%

\author{Tommaso Tufarelli}
\affiliation{%
School of Mathematical Sciences and Centre for the Mathematics and Theoretical Physics of Quantum Non-Equilibrium Systems, University of Nottingham, University Park, Nottingham NG7 2RD, UK}%

\author{Gerardo Adesso}
\email{gerardo.adesso@nottingham.ac.uk}
\affiliation{%
School of Mathematical Sciences and Centre for the Mathematics and Theoretical Physics of Quantum Non-Equilibrium Systems, University of Nottingham, University Park, Nottingham NG7 2RD, UK}%

\author{Samanta Piano}
\email{samanta.piano@nottingham.ac.uk}
\affiliation{%
 {Manufacturing~Metrology~Team,~Faculty~of~Engineering,~University~of~Nottingham,~Nottingham~NG7~2RD,~UK}
}%

\date{\today}

\begin{abstract}
We develop a quantum statistical framework for passive optical surface metrology. Modelling a surface as an incoherent ensemble of point emitters imaged through a diffraction-limited system, we employ techniques from quantum parameter estimation and hypothesis testing to derive ultimate bounds for jointly estimating geometrical features and for deciding the presence or absence of surface defects, and we identify optimal measurements from the geometry of the point-spread-function manifold.  As a representative application, we analyse a minimal surface crack model based on three point sources and show that spatial mode sorting can simultaneously enable near-quantum-limited estimation of crack width and depth and markedly enhanced detectability of the crack, compared with direct imaging. Our results pave the way towards enhanced optical inspection and characterisation of sub-diffraction surface features by probing a limited number of spatial modes without any illumination control.
\end{abstract}

\maketitle

\paragraph{\bfseries Introduction.\label{sec:intro}}
Optical surface metrology enables non-contact measurement of three-dimensional (3D) surface topography using light \cite{leach2020advances,catalucci2022optical}. Compared with contact-based techniques, optical approaches can enable faster and less invasive measurements over extended surface areas \cite{hocken2005optical}. In practice, optical surface metrology encompasses a range of techniques that may rely on controlled illumination ({\em active} methods), such as structured-light or interferometric profilometry \cite{yang2022single,van2019deep,wang2021single,li2016single}, or instead exploit naturally available or uncontrolled illumination ({\em passive} methods) \cite{mathia2011recent,marrugo2020state}. In both cases, the achievable precision is ultimately limited by diffraction and by the information that can be extracted from the detected optical field.

In this Letter, we analyse optical surface metrology from the perspective of {\em quantum} parameter estimation and hypothesis testing, to formulate the ultimate precision bounds and propose measurements approaching them. This line of inquiry originates from Tsang's seminal work \cite{tsang2016quantum}, which demonstrated that quantum measurements can overcome classical resolution limits in estimating the separation between two sources. 
Adopting a non-orthogonal basis formulation \cite{fiderer2021general,napoli2019towards,genoni2019non}, our study reveals that the quantum limits are governed entirely by pairwise overlaps of the  point-spread functions (PSFs) and by the tangent directions generated by infinitesimal displacements. This structure directly identifies physically meaningful measurement modes tailored to specific surface parameters.

We illustrate our framework on the technologically relevant task of {\em estimating the depth and width of a surface crack}. We focus on a minimal model in which the crack profile is represented by three point sources, allowing a complete quantitative analysis based on Fisher information. We additionally address the complementary problem of {\em crack detection}, formulating it as a binary hypothesis test \cite{GG} and deriving analytical and numerical bounds on the Chernoff information governing the discrimination performance. In both estimation and detection settings, we compare classical direct imaging with experimentally accessible {\em spatial mode sorting} \cite{modes,starlight,da2021super,MPLC,Paur:16,Yang:16,Tham:17,zhou2019quantum,Zanfo,Frank:23}. We show that the latter enables substantial precision gains, reaching the ultimate quantum limits using purely passive optical measurements.

\paragraph{\bfseries Preliminaries.\label{sec:prelim}}
We recall the quantum statistical tools that will be used to derive error bounds for surface characterisation.

\paragraph{Parameter estimation theory.\label{sec:parameterestimation}}
We consider $m$ distinct parameters $\theta = (\theta_1,\theta_2,\dots,\theta_m)$ for which we aim to build estimators $ \hat{\theta} = (\hat{\theta}_1,\hat{\theta}_2,\dots,\hat{\theta}_m)$. For any unbiased estimation strategy specified by a measurement $\Pi$, the error covariance matrix $\text{cov}(\theta)$ is lower bounded by the classical Cram\'er-Rao bound \cite{lehmann2006theory,van2004detection}, which in turn is lower bounded by the quantum Cram\'er-Rao bound \cite{helstrom1969quantum,braunstein1994statistical}
\begin{align}
    \text{cov}(\theta) &\geq N^{-1} J_\Pi^{-1} \geq N^{-1} H^{-1},
\end{align}
with $N$ being the number of measurements, $J_\Pi$ the Fisher information matrix (FIM) and $H$ the quantum Fisher information matrix (QFIM). The inequalities mean that $N\text{cov}(\theta) - J_\Pi^{-1}$  and $H-J_\Pi$ are positive semidefinite matrices for all $\Pi$. If we implicitly define the symmetric logarithmic derivative (SLD) $\Lambda_a$ related to the parameter $\theta_a$ ($a=1,\dots,m$) as
\begin{align}
   2{\partial_{\theta_a} \rho} = \rho \Lambda_a+ \Lambda_a \rho,
\label{def:sld}
\end{align}
for a parameter-dependent density operator $\rho=\rho(\theta)$, we can express the QFIM matrix elements as
\begin{align}
    H_{a,b} = \text{Tr} \left [ \Lambda_{a} {\partial_{\theta_b} \rho} \right ].
\label{def:QFIM}
\end{align}
Assuming Poissonian counting statistics \cite{tsang2018subdiffraction} to account for typical optical photon losses, the FIM can be written as  
\begin{align}
    ({J_\Pi})_{a,b} &= \tau \sum_j^M\frac{1}{P_j(\theta)} \pdv{P_j(\theta)}{\theta_a}  \pdv{P_j(\theta)}{\theta_b},
\label{eq:FI_MS}
\end{align}
where $P_j(\theta) = \int dr\, \big|\!\braket{\phi_j}{\psi}\!\big|^2 F(r|\theta)$ is the mean detected power in a measurement channel $\phi_j$ for an object distribution $F(r|\theta)$, with $\ket{\psi}$ the PSF of the optical system, $\tau$ the Poissonian detection rate, and $M$ the number of measurement channels.

If we are interested in estimating a different set of parameters, say $\Tilde{\theta}$, which can be expressed as functions of the original parameters $\theta$, then the corresponding QFIM and FIM are transformed via the Jacobian of the re-parametrisation \cite{paris2009quantum,liu2020quantum},
\begin{align}
    \Tilde{H} &= B H B^\top, \quad
    B_{\alpha,b} = \partial{\theta_b}/\partial{\Tilde{\theta}_\alpha}, \quad     \Tilde{J_\Pi} = B J_\Pi B^\top.
    \label{def:jacobian}
\end{align}

To preserve the physical and geometric picture of the underlying problem, we will use a non-orthogonal basis to calculate the QFIM \cite{genoni2019non,napoli2019towards,fiderer2021general}. Given a (generally non-orthogonal) basis $\{\ket{\psi_v}\}, v = 1,2,\dots$, which spans the subspace relevant to the estimation problem, we can then expand all operators of interest as:
$\rho = \sum_{u,v} R^{u,v} \ket{\psi_u}\bra{\psi_v}$, $\pdv{\rho}{\theta_a} = \sum_{u,v} D_a^{u,v} \ket{\psi_u}\bra{\psi_v}$,
$\Lambda_a =  \sum_{u,v} L_a^{u,v} \ket{\psi_u}\bra{\psi_v}$,
which implicitly define the matrices ${R}, D_a$ and $L_a$. Then, the $a^{\textrm{th}}$ SLD in Eq.~\eqref{def:sld} becomes
\begin{align}
    2 {D}_a = {L}_a{G}{R} + {R}{G}{L}_a,
    \label{eq:sld_nonorthogonal}
\end{align}
and the QFIM Eq.~\eqref{def:QFIM} transforms into
\begin{align}
    H_{a,b} = \Re \Tr (L_a G D_b G) =  \Re \Tr( R G L_a  G L_b G),
    \label{eq:qfi_nonorthogonal}
\end{align}
where we define the Gram matrix $G_{u,v}=\braket{\psi_u}{\psi_v}$ to account for the potential non-orthogonal basis representation. 

\paragraph{Quantum hypothesis testing.\label{sec:hypothesistesting}}
We consider a binary quantum hypothesis test to discriminate between two states $\rho_0$ (null hypothesis) and $\rho_1$ (alternative hypothesis). For a  measurement $\Pi$, a decision rule can yield a type I error (false positive) with probability $\alpha$ and a type II error (false negative) with probability $\beta$. In the asymptotic limit of $N \gg 1$ identical runs, the symmetric error probability  $P_{\textrm{err}} = (\alpha + \beta)/2$ decreases exponentially as $P_{\textrm{err}}(N) \sim e^{-N \xi_{\Pi}}$. The asymptotic error exponent $\xi_{\Pi}$, which quantifies the distinguishability between the two hypotheses, is given by the Chernoff information \cite{cover1999elements},
\begin{equation} \label{eq:CCB}
    \xi_{\Pi} = - \min_{0 \leq s \leq 1} \log \Big( \sum_x P_0(x)^s P_1(x)^{1-s} \Big),
\end{equation}
with $P_j$ the outcome probability distribution for $\rho_j$ ($j=0,1$). Maximizing $\xi_{\Pi}$ over all possible measurements yields the quantum Chernoff bound (QCB) $\xi_Q \geq \xi_\Pi$, which represents the ultimate upper limit on asymptotic distinguishability and can be evaluated directly from the density matrices \cite{audenaert2007discriminating},
\begin{equation} \label{eq:QCB}
\xi_Q = - \min_{0 \leq s \leq 1} \log \text{Tr} \big[ \rho_0^s \rho_1^{1-s} \big].
\end{equation}

\paragraph{\bfseries Quantum precision bounds for discrete imaging models.\label{sec:CR-bounds} }
Assuming no \textit{a priori} control  over the illumination, we model a surface as a discrete grid of points in 3D space, each emitting photons that get collected by a lens. For monochromatic light of wavelength $\lambda$ and a circular aperture, the spatial probability distribution of photons from a single point source in the image plane defines the PSF, described by a Gaussian beam
\begin{equation}
\begin{aligned}
\label{def:psf}
\!\!    &\psi_v (x+\delta x_v,y+\delta y_v,z+\delta z_v) = \sqrt{\frac{1}{\pi}} \frac{i}{z+\delta z_v + i} \\
\!\!    &\times \exp{\frac{-i((x+\delta x_v)^2 + (y+ \delta y_v)^2)}{2(z+\delta z_v + i)} - ikz_R(z+\delta z_v)},
\end{aligned}
\end{equation}
where $(x,y,z)$ are the coordinates in the object plane, $(\delta x_v,\delta y_v,\delta z_v)$ are displacements of the point source $v$ from the origin in the object plane, $k= 2\pi / \lambda$ is the wave vector and $z_R$ the Rayleigh length. We renormalised the coordinate system to use relative length scales, $x \rightarrow \sqrt{k/z_R}x$ and $ z \rightarrow z/z_R$.

To calculate the QFIM Eq.~\eqref{eq:qfi_nonorthogonal}, we can represent an object consisting of $V$ incoherent point sources by the mixed state
\begin{align} \label{eq:incoherent_ensemble}
    R = {\sum}_v^V I_v \ket{\psi_v}\bra{\psi_v},
\end{align}
where $\ket{\psi_v} = \int \psi_v (x+\delta x_v,y+\delta y_v,z+\delta z_v) \ket{xyz}$ is the PSF centered at the displaced point source $(\delta x_v,\delta y_v,\delta z_v)$ from the origin and $I_v$ its relative intensity, with $\sum_v I_v=1$. 

The states ${\ket{\psi_v}}$ are non-orthogonal: their mutual overlaps encode the diffraction-induced indistinguishability between nearby surface features. 
The relevant information about the displacement parameters $\theta = (\theta_v^a) \equiv (\delta x_v,\delta y_v,\delta z_v)$ is contained in the tangent directions generated by infinitesimal shifts of each PSF. 
Since in general $\ket{\partial_{\theta_u^a}\psi_u}$ is linearly independent from the states $\ket{\psi_v}$,
we must  enlarge the support of the Hilbert space to include the tangent subspace spanned by ${\ket{\partial_{\theta_v^a}\psi_v}}$. This extends the representation from dimension $V$ to $V+3V$.
In the following, we will assume all point sources have equal intensity $I_v=1/V$ \cite{noteintensity}, leading to a  $3V \times 3V$ QFIM encoding precision bounds for estimating only the  displacements $\theta_v^a$ for each point source $v=1,\ldots,V$.

Ordering the basis vectors as $\big\{|{\psi}\rangle$, $|{\partial_{\delta x_v} \psi}\rangle$, $|{\partial_{\delta y_v} \psi}\rangle$, $|{\partial_{\delta z_v} \psi}\rangle\big\}$, we write all matrices in block matrix form
\begin{align}\label{eatlean}
    X &= \begin{pmatrix}
        X_\Delta & X_\gamma\\
        X_{\gamma^\dagger} & X_\tau
    \end{pmatrix},
\end{align}
where $X_\Delta \in V \times V$ represents the PSF manifold geometry, $X_\tau \in 3V \times 3V$ the curvature structure (overlaps between tangent vectors), and $X_\gamma \in 3V \times V$ and its adjoint $X_{\gamma^\dagger}$ the first-order sensitivity (PSF–tangent couplings).
Assuming Gaussian beams Eq.~\eqref{def:psf}, the Gram matrix $G$ can be evaluated analytically (see Appendix~\ref{app:gaussian_overlaps} \cite{supposta}),
while the density matrix has support only on the first block $\Delta$, $R_\Delta = \frac{1}{V} \openone_V$,
with $\openone_V$ the $V\times V$ identity matrix, and each derivative matrix becomes sparse with only two non-zero entries, 
    $D_a^{\partial u,u} = D_a^{u,\partial u} = \frac{1}{V}$ (where the $\partial u$ superscript corresponds to the basis vector $\ket{\partial_{\theta_u^a} \psi_u}$).
Solving for the QFIM in this basis results in the closed-form solution (see Appendix~\ref{app:qfi_derivation} \cite{supposta})
\begin{align}
    H_{a,b} &= 2 \Re \langle{\gamma^b}| G_{\gamma^\dagger} L^a_\Delta G_\Delta |{\gamma^b}\rangle \nonumber\\
    &+ 4 \Re \langle{\gamma^a}|G_{\gamma^\dagger}|{\gamma^b}\rangle\langle{\gamma^b}|G_{\gamma^\dagger} G^{-1}_\Delta|{\gamma^a}\rangle,
    \label{eq:qfim_final}
\end{align}
where $G^{-1}_\Delta$ is the inverse PSF overlap matrix, $\ket{\gamma^a}$ (resp. $\ket{\gamma^b}$) are the basis vectors encoding derivative matrices $D_{\gamma^{a}}$ (resp. $D_{\gamma^{b}}$), and $L^a_\Delta$ is defined by the reduced Lyapunov equation
\begin{align}
    L^a_{\Delta} G_{\Delta} + G_{\Delta} L^a_{\Delta}  &= -2 (G_{\gamma}D_{\gamma^{a\dagger}}G_\Delta^{-1} + G_\Delta^{-1} D_{\gamma^a}G_{\gamma^\dagger}).
    \label{eq:L_delta_sylvester}
\end{align}
Since $G_\Delta$ is Hermitian and invertible, each $L_\Delta$ is unique and the corresponding SLD operators are Hermitian. The final expression for the QFIM Eq.~\eqref{eq:qfim_final} shows that the quantum Fisher information depends solely on the PSF overlap matrix $G_\Delta$ and its couplings to first-order derivatives  $G_{\gamma} $ (resp. $G_{\gamma^\dagger}$). 
The ultimate precision bounds are therefore determined entirely by the local overlap geometry of the PSF manifold. For a Gaussian PSF, the derivative directions are proportional to Hermite-Gaussian functions of increasing order, which motivates the study of spatial mode sorting in the Hermite polynomial basis \cite{weisstein2002hermite} as an optimal measurement strategy \cite{starlight,rehacek2017optimal}.

\begin{figure}[t]
\includegraphics[width=.6\columnwidth]{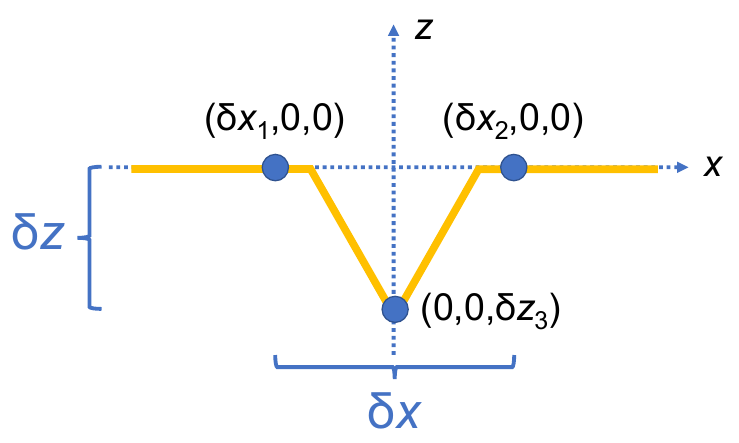} \vspace*{-0.25cm}
\caption{\label{fig:3source_layout}Representation of a surface crack with width parameter $\delta x$ and depth parameter $\delta z$, modelled by three incoherent point sources.}
\end{figure}

\paragraph{\bfseries Application: Surface crack inspection.\label{sec:3src}}
We now apply our general methods to the concrete problem of analysing {\em surface cracks}. We will explore two tasks: (i) estimating geometric parameters of a crack, via Fisher information, and (ii) detecting the presence of a crack, via Chernoff information.

In our toy model, a crack is described by three point sources placed symmetrically on the $x$-axis, with the third (middle) one displaced along the $z$-axis [Fig.~\ref{fig:3source_layout}]. We relate the coordinates of the point sources to two parameters that characterise the problem, i.e., the width $\delta x$ and depth $\delta z$ of the crack,
\begin{equation}
\begin{aligned}
    \delta x &\coloneqq {\delta x_2 - \delta x_1}, \quad     \delta z \coloneqq -\delta z_3.\
\end{aligned}
\label{eq:3pt2paramcoords}
\end{equation}

\paragraph{(i) Estimating width and depth of the crack.}
The ultimate precision bound on estimating individual source displacements (${\delta x_1},{\delta x_2},{\delta z_3}$) is given by the QFIM from Eq.~(\ref{eq:qfim_final}). We collapse the QFIM to a $2 \times 2$ matrix for the parameters of interest $\delta x$ and $\delta z$  using Eq.~(\ref{def:jacobian}) with the Jacobian matrix $B = \left(\begin{smallmatrix}
    -\frac{1}{2} & \frac{1}{2} &  0\\
    0 &  0 & -1
    \end{smallmatrix}\right)$.

We now compare two measurement strategies against the quantum precision bound, namely quantum-inspired mode sorting (MS) in the Hermite polynomial basis and the classical direct imaging (DI) approach most commonly utilised. For the MS basis we match the Hermite-Gaussian modes to the PSF, $\phi_{j,l}(x,y) = \frac{H_j(x)}{\sqrt{2^j j!}} \frac{H_l(y)}{\sqrt{2^l l!}} \sqrt{\frac{1}{\pi}} \exp{-\frac{(x^2 + y^2)}{2}}$ and compute the FIM via Eq.~\eqref{eq:FI_MS} using measurement channel probabilities
\begin{align}
    P_{j,l}(\theta) &= \frac{1}{V} \sum_v^V  \abs{ \braket{\phi_{j,l}}{\psi_v(x + \delta x_v, y + \delta y_v, + \delta z_v)} }^2 \label{eq:prob_ms_general}
\end{align}
with closed-form expressions provided in Appendix~\ref{app:ms_incoherent} \cite{supposta}. For DI, the measurement distribution in the image plane is a sum of the individual intensities $P(x,y|\delta x,\delta z)=\frac13\left[{|{\psi_1(x\!-\!\frac{\delta x}{2},y,0)}|^2\!\! + \!|{\psi_2(x\!+\!\frac{\delta x}{2},y,0)}|^2\!\! + \!|{\psi_3(x,y,-\delta z)}|^2}\right]$, which we integrate numerically to obtain the classical FIM.

We find (see Fig.~\ref{fig:supp4figs} in \cite{supposta}) that the Fisher information confirms the geometric intuition from the QFIM. For estimating the crack width in the sub-Rayleigh regime ($\delta x \rightarrow 0$), the information concentrates  in the $(1,0)$ Hermite-Gaussian mode, corresponding to the first-order transverse PSF derivative. At larger separations, the information spreads across higher-order modes limiting the achievable precision, while  DI approaches the quantum limit instead. Similarly, for depth estimation  ($\delta z\rightarrow0$), MS approaches the quantum limit with equal dominant contributions from the $(2,0)$ and $(0,2)$ modes, which are proportional to second-order transverse PSF derivatives  and span the tangent directions associated with axial defocus. In contrast, the DI Fisher information peaks at the Rayleigh length, consistent with classical beam propagation.

\begin{figure}
\centering
\begin{overpic}[height=4cm]{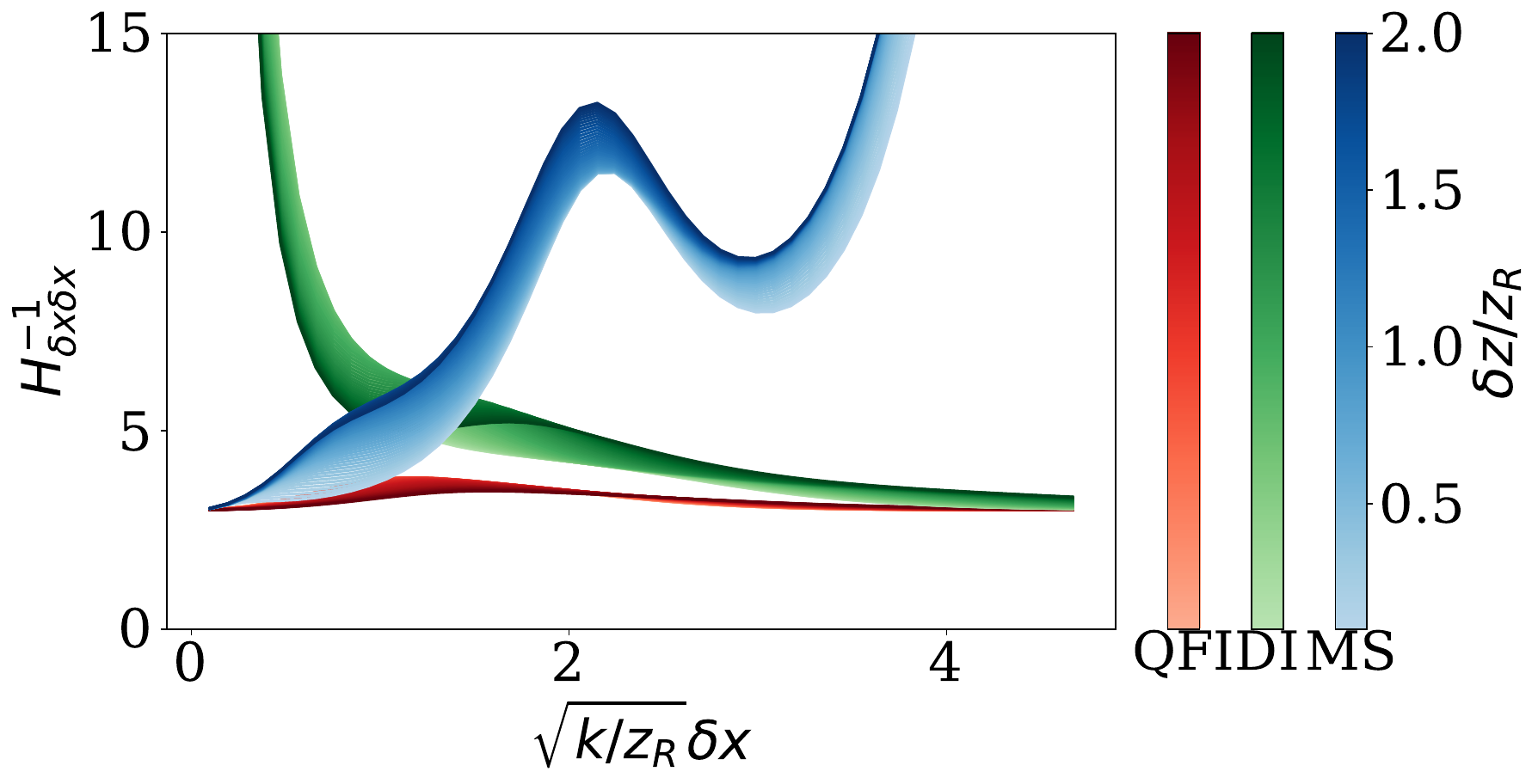}
    \put(1, 2){{(a)}} 
\end{overpic} \\
\vspace{0.2cm} 
\begin{overpic}[height=4cm]{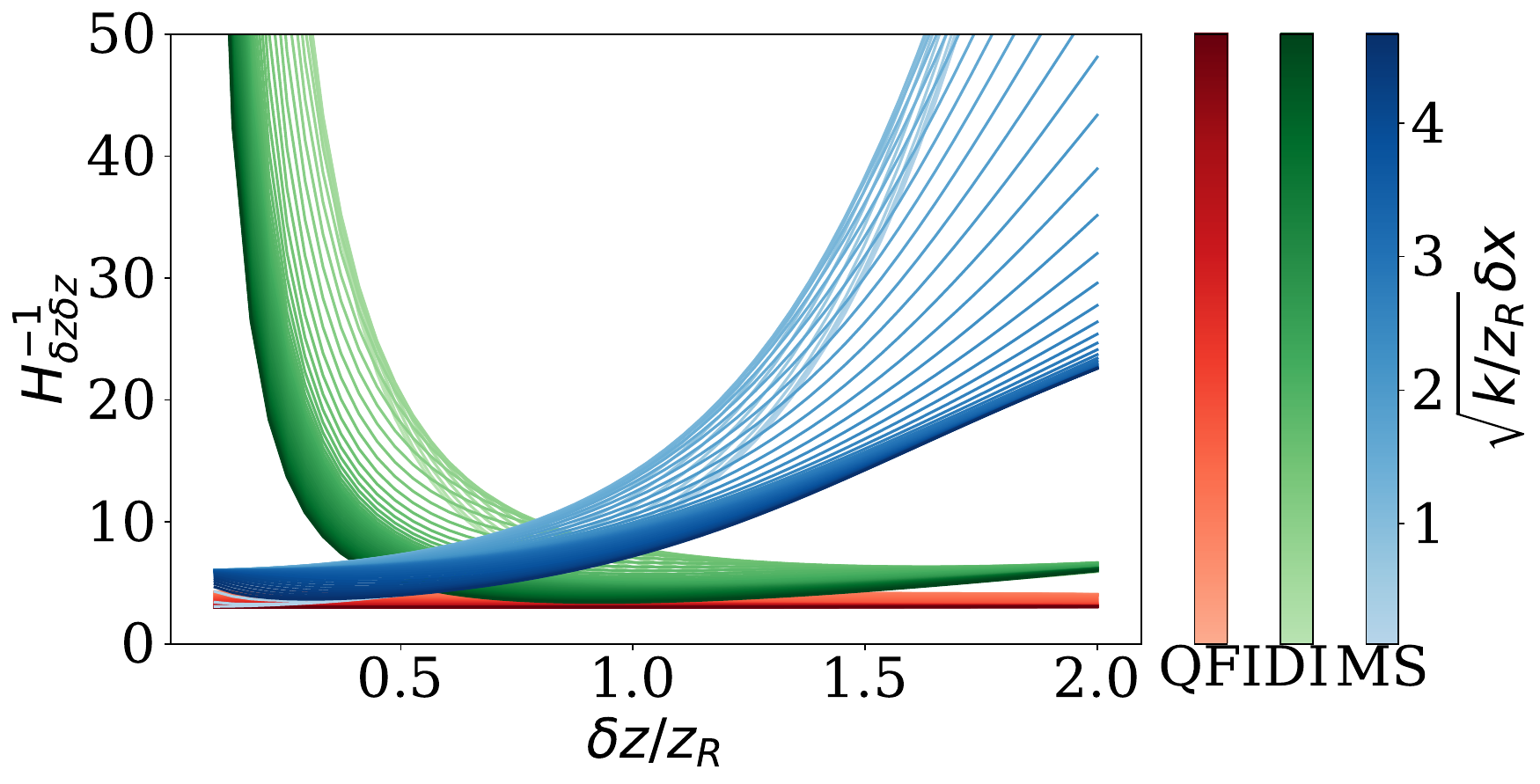}
    \put(1, 2){{(b)}}
\end{overpic}
\caption{Cram\'er-Rao bounds for estimating surface crack parameters (a) $\delta x$ and (b) $\delta z$ as modelled in Fig.~\ref{fig:3source_layout}. Red curves correspond to the ultimate quantum limit (QFI), blue curves to Hermite-Gaussian mode sorting (MS), and green curves to classical direct imaging (DI).}
\label{figshers}
\end{figure}

To obtain the Cram\'er-Rao bounds we need to invert the multiparameter FIM. For finite $\delta x$ and $\delta z$, the off-diagonal elements of the FIM are non-zero, showing that width and depth are statistically correlated under generic measurements \cite{ragy2016compatibility}. Nevertheless, measuring in the Hermite-Gaussian mode basis shows the Cram\'er-Rao bound approaching the quantum limit for both $\delta x$ [Fig.~\ref{figshers}(a)] and $\delta z$ [Fig.~\ref{figshers}(b)]. This shows that simultaneous sub-Rayleigh estimation of the crack width and depth is possible using appropriate quantum-inspired measurements.

\paragraph{(ii) Detecting crack versus no crack.}
We now show that the techniques presented so far are also advantageous for deciding whether the object we are imaging is a crack or not. This decision problem can be characterised as a quantum hypothesis test between the images of a crack (with given width $\delta x$ and depth $\delta z$, see Fig.~\ref{fig:3source_layout}) and its absence. Let $\psi_L \equiv \psi_1$ and $\psi_R \equiv \psi_2$ be sources on the edges at $(\pm \delta x/2,0,0)$,   $\psi_C$ be on their midpoint at $(0,0,0)$ representing the centre of a flat surface, and $\psi_Z \equiv \psi_3$ be on the floor of a crack at $(0,0,-\delta z)$.
We can then define the states corresponding to each of our hypotheses.
For the null hypothesis, $\rho_0$ describes three collinear sources representing a flat surface (no crack); whereas for the alternative hypothesis, $\rho_1$ denotes a crack with depth $\delta z$,
\begin{equation}
    \begin{aligned}
        \rho_0 &= \left({\ketbra{\psi_L}{\psi_L} + \ketbra{\psi_R}{\psi_R} + \ketbra{\psi_C}{\psi_C}}\right)\!/3\,,\\
        \rho_1 &= \left({\ketbra{\psi_L}{\psi_L} + \ketbra{\psi_R}{\psi_R} + \ketbra{\psi_Z}{\psi_Z}}\right)\!/3\,.
    \end{aligned}
\end{equation}
We can now apply the definitions in Eqs.~(\ref{eq:CCB}) and (\ref{eq:QCB}) to calculate the QCB and the  Chernoff information of each measurement.

\begin{figure}
\centering
\begin{overpic}[width=7cm]{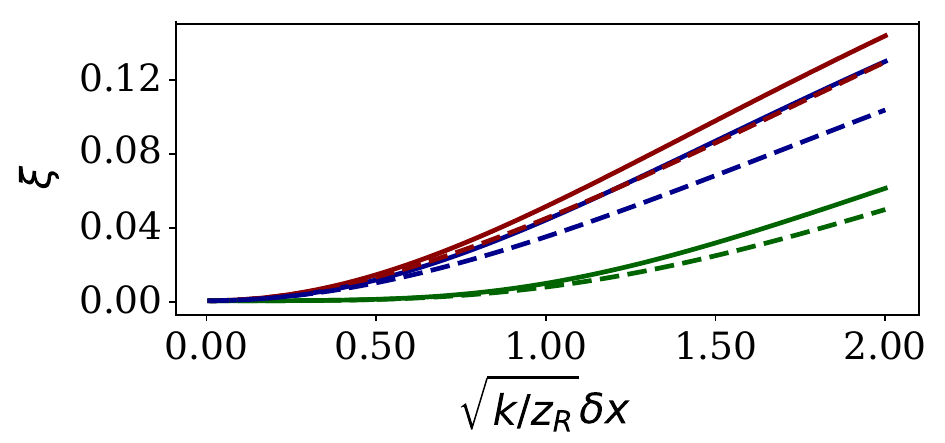}
    \put(1, 3){{(a)}} 
\end{overpic} \\
\vspace{0.2cm} 
\begin{overpic}[width=7cm]{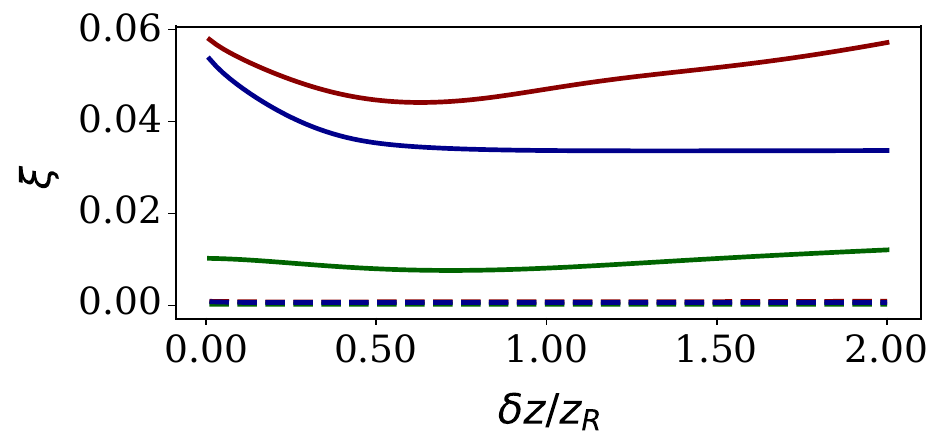}
    \put(1, 3){{(b)}}
\end{overpic}
\caption{Chernoff information for detecting surface cracks, plotted as a function of: (a) the width $\delta x$ for $\delta z/z_R=1$ (solid) and $0.1$ (dashed); (b) the depth $\delta z$ for $\sqrt{k/z_R}\delta_x=2$ (solid) and $0.5$ (dashed). Red curves correspond to the quantum Chernoff bound, blue curves to Hermite-Gaussian mode sorting, and green ones to direct imaging.}
\label{chernoffigo}
\end{figure}

To calculate the QCB, we take two approaches. First, a numerical approach (See Appendix~\ref{app:QCB_calculation} \cite{supposta}) with  results shown in Fig.~\ref{chernoffigo}. Second, an analytical upper bound using \cite{calsamiglia2008quantum,lupo2020quantum}
\begin{equation} \label{eq:fidelity_bound}
    \xi_Q \leq - 2 \log F(\rho_0, \rho_1),
\end{equation}
where $F(\rho_0, \rho_1) = \text{Tr}\left[ \sqrt{\sqrt{\rho_0} \rho_1 \sqrt{\rho_0}}\right]$ is the fidelity between $\rho_0$ and $\rho_1$, which we express in terms of point source overlaps in Appendix~\ref{app:QCB_calculation} \cite{supposta}. By using leading-order approximations, valid for small $\delta x$ and $\delta z$, we obtain $F(\rho_0,\rho_1) \approx 1 - \frac{\delta z^2}{24}$. We can then use Eq.~\eqref{eq:fidelity_bound} to provide a bound on the QCB,
\begin{equation}
    \xi_Q \leq \frac{\delta z^2}{12} + O(\delta z^4).
\end{equation}

We now 
turn our attention to calculating the Chernoff information Eq.~(\ref{eq:CCB}) for a classical DI intensity measurement. To do this, we first use the sum of the intensity distributions of each point source to obtain a probability distribution for each hypothesis: $P_0(x,y) = \big({|\psi_L|^2 + |\psi_R|^2 + |\psi_C|^2}\big)/{3}$ and $P_1(x,y) = \big({|\psi_L|^2 + |\psi_R|^2 + |\psi_Z|^2}\big)/{3}$. The Chernoff information for DI can then be calculated from Eq.~(\ref{eq:CCB}):
\begin{equation}\label{eq:CI_DI}
\xi_{\textrm{DI}} = - \log \min_{0 \leq s \leq 1} \mbox{$\iint P_0(x,y)^s P_1(x,y)^{1-s} dx dy$}.
\end{equation}
Like the QCB, we use numerical methods to evaluate this quantity as well as a series expansion of the integrand which is then integrated term by term to give us the leading term of the error exponent: $\xi_{\textrm{DI}} \approx - \log \min\limits_{0 \leq s \leq 1} \big( 1 - s(1-s) \frac{\delta z^4}{18}\big) = -\log \big(1 - \frac{\delta z^4}{72} \big)$.
Using the approximation $-\log(1-\epsilon) \approx \epsilon$ for $\epsilon \ll 1$, we then get the approximation $\xi_{\textrm{DI}} \approx \frac{\delta z^4}{72}$. This quartic scaling implies a severe loss of distinguishability for shallow cracks by DI compared to the quantum limit.

To benchmark MS measurements, we define $P_{j,l}(\delta x, \delta z)$ as the combined probability of detecting channel $(j,l)$ from all three point sources in a crack with given width $\delta x$ and depth $\delta z$, using Eq.~\eqref{eq:prob_ms_general}. To ensure valid probability distributions, we implicitly include a remainder channel with probability $1 - \sum_{j,l} P_{j,l}(\delta x, \delta z)$.
Noting that the null hypothesis is represented by $P_{j,l}(\delta x, 0)$, 
the MS Chernoff information reads 
\begin{equation} \label{eq:CI_MS}
\xi_{\textrm{MS}} = -\log \min_{0 \leq s \leq 1} \mbox{$\sum_{j,l} P_{j,l}(\delta x, 0)^s P_{j,l}(\delta x, \delta z)^{1-s}$}.
\end{equation}
As before, for MS in the Hermite-Gaussian basis we calculate this quantity numerically and give a leading-order approximation.
The approximation is obtained by noting that only four modes contain terms of second order:
		$P_{0,0}(\delta x, \delta z) \approx 1 - \frac{\delta x^2}{12} - \frac{\delta z^2}{12}$,
        $P_{1,0}(\delta x, \delta z) \approx \frac{\delta x^2}{12}$,
		$P_{2,0}(\delta x, \delta z) \approx \frac{\delta z^2}{24}$,
		$P_{0,2}(\delta x, \delta z) \approx \frac{\delta z^2}{24}$.
The channels $(2,0)$ and $(0,2)$ have particular significance as the leading-order term vanishes for a flat surface ($P_{2,0}(\delta x, 0) \approx P_{0,2}(\delta x, 0) \approx 0$) but have nonzero probability $\propto \delta z^2$ for a crack, resulting in  increased distinguishability.
By substituting our approximation into Eq.~(\ref{eq:CI_MS}), we get $\xi_{\textrm{MS}}  \approx -\log \min\limits_{0 \leq s \leq 1} \big(1  - \frac{\delta z^2}{12} + \frac{\delta z^2 s}{12} \big) \approx \frac{\delta z^2}{12}$. Since our leading order bound gives $\xi_Q \leq \frac{\delta z^2}{12}$ and $\xi_{\textrm{MS}}  \approx \frac{\delta z^2}{12}$ matches this bound, we conclude that $\xi_Q \approx \frac{\delta z^2}{12}$.

 Our hypothesis testing results, summarised in Table~\ref{tab:chernoff_comparison}, show that MS saturates the quantum limit up to second-order scaling with $\delta z$, thanks to the activation of specific channels sensitive to the crack depth. In contrast, DI produces only a small change to the intensity distribution, making shallow cracks much more difficult to distinguish from background noise.

\begin{table}[h]
\centering
\begin{tabular}{lc}
\hline \hline
\  Scheme & \qquad  Chernoff exponent \qquad  \\ \hline 
\  Quantum limit (QCB) &\qquad  ${\delta z^2}/{12}$\qquad \\
\  Direct imaging (DI) &\qquad ${\delta z^4}/{72}$ \qquad\\
\  Mode sorting (MS) & \qquad ${\delta z^2}/{12}$\qquad\\ \hline \hline
\end{tabular}
\caption{Comparison of leading-order scaling of Chernoff exponents for detection of a surface crack with depth $\delta z$.}
\label{tab:chernoff_comparison}
\end{table}

\paragraph{\bfseries Conclusions.}
We have introduced a quantum statistical framework for passive optical surface metrology that unifies fundamental limits for parameter estimation and defect detection within a single description. By modelling a surface as an incoherent ensemble of point emitters imaged through a diffraction-limited system, we connected the estimability and detectability of physically meaningful surface parameters directly to Fisher information and Chernoff information bounds.
These bounds depend solely on point-spread function overlaps and their derivative structure, providing a streamlined geometric interpretation for the design of optimal measurements.

Applied to a minimal surface crack model, our analysis shows that specific spatial modes are selectively sensitive to distinct geometric features, such as crack width and depth. This enables simultaneous near-quantum-limited estimation of multiple parameters using only a small number of low-order modes, despite the absence of illumination control. Beyond estimation, the same measurement strategy markedly enhances the detection of shallow defects at the ultimate quantum limit, while classical direct imaging performs substantially worse.

Taken together, these results demonstrate that quantum-limited performance in surface metrology is, in principle, achievable with purely passive measurements in the image plane. More broadly, they show how quantum estimation and detection theory can be used not only to bound performance, but also to identify physically interpretable measurement bases tailored to specific quality control and surface analysis tasks.

\begin{acknowledgments}

We are grateful to M.~Guta, H.~Hooshmand, A.~Khan, A.~Lvovsky, N.~Treps, and S.~da~Vinci for fruitful exchanges.   
We acknowledge financial support from BBSRC (Grant No.~BB/X004317/1) and EPSRC (Grants No.~EP/X010929/1 and EP/Z534948/1).
\end{acknowledgments}

\bibliography{ref}

\clearpage
\appendix
\onecolumngrid
\setcounter{secnumdepth}{2}
\setcounter{figure}{0}
\renewcommand{\thefigure}{S\arabic{figure}}

\section*{SUPPLEMENTAL MATERIAL}

\section{Closed form Gaussian overlaps\label{app:gaussian_overlaps}}
Here we show the explicit closed form expressions of the overlaps entering the Gram matrix $G$ for a Gaussian beam,
\begin{equation}
    G_{\Delta_{v,u}} = \braket{\psi_v}{\psi_u}, \qquad    G_{\gamma^{a}_{v,u}} = \braket{\psi_v}{ \frac{\partial \psi_u}{\partial \theta_u^a}}, \qquad G_{\tau^{a,b}_{v,u}} = \braket{\frac{\partial \psi_v}{\partial \theta_v^a}}{ \frac{\partial \psi_u}{\partial \theta_u^b}}.
    \label{eq:Grammian}
\end{equation}

All overlaps share a common exponential factor
\begin{align}
    \psi_0 = \exp \left \{ (i \frac{(\delta_{x_v} - \delta_{x_u})^2 + (\delta_{y_v} - \delta_{y_u})^2)}{2(-2i + \delta_{z_v} - \delta_{z_u})} + i k z_R (\delta_{z_v} - \delta_{z_u}) \right \}.
\end{align}

We have then
\begin{align}
    \braket{\psi_v}{\psi_u} &= \frac{2 i}{2 i  - \delta_{z_v} + \delta_{z_u}} \psi_0 \\
    \braket{\partial_{\delta_{x_v}}\psi_v}{\psi_u} &=  -2 \frac{\delta_{x_v} - \delta_{x_u}}{(2i - \delta_{z_v} + \delta_{z_u})^2} \psi_0\\
    \braket{\partial_{\delta_{y_v}}\psi_v}{\psi_u} &=  -2 \frac{\delta_{y_v} - \delta_{y_u}}{(2i - \delta_{z_v} + \delta_{z_u})^2} \psi_0\\
    \braket{\partial_{\delta_{z_v}}\psi_v}{\psi_u} &= \frac{-4 + (\delta_{x_v} - \delta_{x_u})^2 + (\delta_{y_v} - \delta_{y_u})^2 - 2 k z_R (-4 + (\delta_{z_v} - \delta_{z_u})^2) +  2i (\delta_{z_v} - \delta_{z_u})  (-1 + 4 k z_R)
    }{(-2i + \delta_{z_v} - \delta_{z_u})^3} \psi_0 \\
    \braket{\partial_{\delta_{x_v}}\psi_v}{\partial_{\delta_{x_u}}\psi_u} &= \frac{2i (-2 + (\delta_{x_v} - \delta_{x_u})^2 + i (\delta_{z_u} - \delta_{z_v}))}{(2i - \delta_{z_v} + \delta_{z_u})^3} \psi_0 \\
    \braket{\partial_{\delta_{x_v}}\psi_v}{\partial_{\delta_{y_u}}\psi_u} &= \frac{2i (\delta_{x_v} - \delta_{x_u}) (\delta_{y_v} - \delta_{y_u})}{ (2i - \delta_{z_v} + \delta_{z_u})^3} \psi_0 \\
    \braket{\partial_{\delta_{x_v}}\psi_v}{\partial_{\delta_{z_u}}\psi_u} &= -\frac{(-1 + i \delta_{z_u})^3 (\delta_{x_v} - \delta_{x_u}) \psi_0}{(i + \delta_{z_u})^3 (2i - \delta_{z_v} + \delta_{z_u})^4} \\
    &(-8 + (\delta_{x_v} - \delta_{x_u})^2 + (\delta_{y_v} - \delta_{y_u})^2  - 2 k z_R (-4 + (\delta_{z_v} - \delta_{z_u})^2) + 4i (\delta_{z_v} - \delta_{z_u}) (-1 + 2 k z_R)) \nonumber \\
    \braket{\partial_{\delta_{y_v}}\psi_v}{\partial_{\delta_{y_u}}\psi_u} &= 2i \frac{-2 + (\delta_{y_v} - \delta_{y_u})^2 + i (\delta_{z_u} - \delta_{z_v})}{ (2i - \delta_{z_v} + \delta_{z_u})^3} \psi_0\\
    \braket{\partial_{\delta_{y_v}}\psi_i}{\partial_{\delta_{z_u}}\psi_u} &= - \frac{(-1 + i \delta_{z_u})^3 (\delta_{y_v} - \delta_{y_u}) \psi_0}{(i + \delta_{z_u})^3 (2i - \delta_{z_v} + \delta_{z_u})^4}  \\
    & (-8 + (\delta_{x_v} - \delta_{x_u})^2 + (\delta_{y_v} - \delta_{y_u})^2 - 2 k z_R (-4 + (\delta_{z_v} - \delta_{z_u})^2) + 4i (\delta_{z_v} - \delta_{z_u}) (-1 + 2 k z_R)) \nonumber \\
    \braket{\partial_{\delta_{z_v}}\psi_v}{\partial_{\delta_{z_u}}\psi_u} &= \frac{i \psi_0}{2 (-2i + \delta_{z_v} - \delta_{z_u})^5} \bigg(  \\
    & -32 + 8 (\delta_{z_v} - \delta_{z_u})^2 - (\delta_{x_v} - \delta_{x_u})^2 ((\delta_{x_v} - \delta_{x_u})^2 + 2 (-8 + (\delta_{y_v} - \delta_{y_u})^2)) -(\delta_{y_v} - \delta_{y_u})^2 (-16 + (\delta_{y_v} - \delta_{y_u})^2) \nonumber \\
    &  + 64 k z_R + 4 k z_R
        (
            (\delta_{z_v} - \delta_{z_u})^2 (-12 + (\delta_{x_v} - \delta_{x_u})^2 + (\delta_{y_v} - \delta_{y_u})^2)
            - 4 ((\delta_{x_v} - \delta_{x_u})^2 + (\delta_{y_v} - \delta_{y_u})^2)
        ) \nonumber \\
    &- (2 k z_R)^2
        (
            (-4 + 4 \delta_{z_v} + \delta_{z_v}^2 - 2 (2 + \delta_{z_v}) \delta_{z_u} + \delta_{z_u}^2)
            (-4 + \delta_{z_v}^2 - 2 \delta_{z_v} (2 + \delta_{z_u}) + \delta_{z_u} (4 + \delta_{z_u}))
        ) \nonumber \\
    &+ i
        (8 (\delta_{z_v} - \delta_{z_u}) ( -4 + (\delta_{x_v} - \delta_{x_u})^2 + (\delta_{y_v} - \delta_{y_u})^2 - k z_R (-12 + (\delta_{z_v} - \delta_{z_u})^2) \nonumber \\
        &- 2 k z_R ((\delta_{x_v} - \delta_{x_u})^2 + (\delta_{y_v} - \delta_{y_u})^2) + (2 k z_R)^2 (-4 + (\delta_{z_v} - \delta_{z_u})^2)))\bigg)\nonumber
\end{align}

\section{Derivation of the Quantum Fisher Information Matrix\label{app:qfi_derivation}}
We will derive the final QFIM expression Eq.~\eqref{eq:qfim_final} in two steps. Firstly, let us show that the elements are fully determined by the Gram matrix Eq.~(\ref{eq:Grammian})
and first-quadrant SLDs Eq.~\eqref{eq:L_delta_sylvester}. Writing the SLD in the same matrix block form as Eq.~(\ref{eatlean}), we can compute the Lyapunov equation Eq.~\eqref{eq:sld_nonorthogonal}
\begin{align}
    2 \begin{pmatrix}
        0 & \Tilde{D}_\gamma \\
        \Tilde{D}_{\gamma^\dagger} & 0
    \end{pmatrix} = \begin{pmatrix}
        \openone_V & 0 \\
        0 & 0
    \end{pmatrix} \begin{pmatrix}
        G_\Delta & G_\gamma \\
        G_{\gamma^\dagger} & G_{\tau}
    \end{pmatrix} \begin{pmatrix}
        L_\Delta & L_\gamma \\
        L_{\gamma^\dagger} & L_{\tau}
    \end{pmatrix} + \begin{pmatrix}
        L_\Delta & L_\gamma \\
        L_{\gamma^\dagger} & L_{\tau}
    \end{pmatrix} \begin{pmatrix}
        G_\Delta & G_\gamma \\
        G_{\gamma^\dagger} & G_{\tau}
    \end{pmatrix} \begin{pmatrix}
        \openone_V & 0 \\
        0 & 0
    \end{pmatrix}, 
\end{align}
where for brevity we omit the parameter index $a$, and we can cancel the $1/V$ factor of the density matrix $V R$ on the right hand side and the derivative matrix $\Tilde{D} = V D$ on the left hand side. Performing the multiplication leads to
\begin{align}
    2 \begin{pmatrix}
        0 & \Tilde{D}_\gamma \\
        \Tilde{D}_{\gamma^\dagger} & 0
    \end{pmatrix} = \begin{pmatrix}
        G_\Delta L_\Delta + L_\Delta G_\Delta + G_\gamma L_{\gamma^\dagger} + L_\gamma G_{\gamma^\dagger} & G_\Delta L_\gamma + G_\gamma L_\tau \\
        L_{\gamma^\dagger} G_\Delta + L_\tau G_{\gamma^\dagger} & 0
    \end{pmatrix}.
\end{align}
Since $R$ only spans support on the first quadrant $\Delta$ subspace, we can without loss of generality set $L_{\tau} = 0$ because the SLD is only defined on the support of $R$, and any component acting outside the support does not affect the QFIM \cite{liu2020quantum}. By element-wise comparison we obtain three matrix equations
\begin{align}
    2 \Tilde{D}_\gamma &= G_\Delta L_\gamma \\
    2 \Tilde{D}_{\gamma^\dagger} &= L_{\gamma^\dagger} G_\Delta \\
    0 & = G_\Delta L_\Delta + L_\Delta G_\Delta + G_\gamma L_{\gamma^\dagger} + L_\gamma G_{\gamma^\dagger}.\label{eq:reduced_sylvester}
\end{align}
Due to our basis construction, the Gram sub-matrix is invertible. Let us write the inverse of $G_\Delta$ as $G^{-1}_\Delta$, then we can formally solve for $L_\gamma$ and $L_{\gamma^\dagger}$ and insert them into Eq.~\eqref{eq:reduced_sylvester} to obtain Eq.~\eqref{eq:L_delta_sylvester}.

Secondly, to derive the QFIM expression we will begin with the definition Eq.~\eqref{eq:qfi_nonorthogonal} and note that due to the properties of computing the trace of a matrix we only need to compute the first quadrant $\Delta$ and fourth quadrant $\tau$, which results in
\begin{align}
    H_{a,b} = \Re \Tr (L_a G D_b G) &= \Re \bigg[\Tr (L^a_\Delta G_\Delta D^b_\gamma G_{\gamma^\dagger}) + \Tr (L^a_\gamma G_{\gamma^\dagger} D^b_\gamma G_{\gamma^\dagger}) + \Tr (L^a_{\Delta} G_{\gamma} D^b_{\gamma^\dagger} G_\Delta) \nonumber \\ &\quad\quad\ +
    \Tr (L^a_{\gamma} G_\tau D^b_{\gamma^\dagger} G_\Delta)  
    + \Tr (L^a_{\gamma^\dagger} G_\Delta D^b_{\gamma} G_\tau)  
    + \Tr (L^a_{\gamma^\dagger} G_{\gamma} D^b_{\gamma^\dagger} G_{\gamma}) \bigg].
\end{align}
We can permute the arguments of a trace, substitute again for $L_\gamma, L_{\gamma^\dagger}$ and write out explicitly the derivative matrices
\begin{align}
    H_{a,b} &= \Re \bigg[
     \Tr (L^a_\Delta G_\Delta \ket{\gamma^b}\bra{\gamma^b} G_{\gamma^\dagger}) 
    +  \Tr (2 G^{-1}_\Delta\ket{\gamma^{a}}\bra{\gamma^{a}} G_{\gamma^\dagger} \ket{\gamma^b}\bra{\gamma^b} G_{\gamma^\dagger}) 
    +  \Tr (L^a_{\Delta} G_{\gamma} \ket{\gamma^{b\dagger}}\bra{\gamma^{b\dagger}} G_\Delta) \nonumber  \\
    &\quad\qquad + \Tr (2 \ket{\gamma^a}\bra{\gamma^a} G_\tau \ket{\gamma^{b\dagger}}\bra{\gamma^{b\dagger}}) 
    +  \Tr (2\ket{\gamma^{a\dagger}}\bra{\gamma^{a\dagger}} G_\Delta \ket{\gamma^b}\bra{\gamma^b}) 
    +  \Tr (2\ket{\gamma^{a\dagger}}\bra{\gamma^{a\dagger}}G^{-1}_\Delta G_{\gamma} \ket{\gamma^{b\dagger}}\bra{\gamma^{b\dagger}} G_{\gamma}) 
 \bigg] \nonumber \\
    &= \Re \bigg[
    \bra{\gamma^b} G_{\gamma^\dagger} L^a_\Delta G_\Delta\ket{\gamma^b} 
    +  2 \bra{\gamma^{a}}G_{\gamma^\dagger}\ket{\gamma^b} \bra{\gamma^b} G_{\gamma^\dagger} G^{-1}_\Delta  \ket{\gamma^{a}} 
\nonumber    \\ & \quad\qquad+  \bra{\gamma^{b\dagger}} G_\Delta L^a_{\Delta} G_{\gamma} \ket{\gamma^{b\dagger}}  + 2\bra{\gamma^{b\dagger}} G_{\gamma} \ket{\gamma^{a\dagger}}\bra{\gamma^{a\dagger}}G^{-1}_\Delta G_{\gamma}\ket{\gamma^{b\dagger}} \bigg]\nonumber \\
    &= 2\Re \bra{\gamma^b} G_{\gamma^\dagger} L^a_\Delta G_\Delta \ket{\gamma^b} 
    + 4 \Re \bra{\gamma^a}G_{\gamma^\dagger}\ket{\gamma^b}\bra{\gamma^b}G_{\gamma^\dagger} G^{-1}_\Delta\ket{\gamma^a},
\end{align}
where in the first step we use the sparsity to reduce the trace to an inner product and in the second step the Hermitian property of all the operators to combine terms.

\section{Classical Fisher Information for an arbitrary polynomial measurement basis\label{app:ms_incoherent}}
To calculate the FIM for a mode measurement Eq.~\eqref{eq:FI_MS}, we first represent the object as an arbitrary discrete distribution of $V$ point sources
\begin{align}
    F(r|\theta) = \frac{1}{V} \sum_v^V \delta \left [ x + \delta x_v, y + \delta y_v, \delta z_v \right ],
\end{align}
with $\delta[x]$ the Dirac delta. Defining the object distribution via a sum of Dirac delta functions simplifies calculating the measurement channel probabilities to Eq.~\eqref{eq:prob_ms_general}.

For each measurement channel contribution $\frac{1}{P_j(\theta)} \pdv{P_j(\theta)}{\theta_a}  \pdv{P_j(\theta)}$ to the FIM Eq.~\eqref{eq:FI_MS} we need to compute and sum the individual measurement mode probabilities for a single source $\abs{ \braket{\phi_{j,l}}{\psi_v(x + \delta x_v, y + \delta y_v, + \delta z_v)} }^2$. However, due to construction, calculating the measurement mode sensitivity for a specific point source displacement only involves computing the single source measurement probability derivative
\begin{align}
 \!\!   \pdv{P_{j,l}(\theta)}{\theta_v^a} &= \frac{1}{V} \pdv{}{\theta_v^a} \abs{\braket{\phi_{j,l}}{\psi(x + \delta x_v, y + \delta y_v, + \delta z_v)}}^2.
 \label{eq:pdv_mode_dot_product}
\end{align}
We assume the modes to be a point-spread function adapted basis \cite{rehacek2017optimal} expressed as \begin{align}
    \phi_{j,l}(x,y) = g_j(x)f_l(y)\psi(x,y,0) = \sum_\zeta^j \sum_\xi^l c^*_\zeta d^*_\xi x^\zeta y^\xi \sqrt{\frac{1}{\pi}} \exp{-\frac{(x^2 + y^2)}{2}},
\end{align}
where $g_j(x)$ (resp.~$f_l(y)$) are polynomials of order $j$ (resp.~$l$) separable in $xy$-dimensions. The measurement overlap between the point source distribution and our measuring mode becomes

\begin{align}
    \braket{\phi_{j,l}}{\psi(x+\delta_x,y+\delta_y,\delta_z)} &= \iint dxdy \  g^*_j(x) \ f^*_l(y) \ \psi(x,y,0) \ \psi(x+\delta_x,y,\delta_z) \nonumber \\
     &= \iint dxdy \ \left [  \sum_\zeta^j \sum_\xi^l c^*_\zeta d^*_\xi x^\zeta y^\xi \sqrt{\frac{1}{\pi}} \exp{-\frac{(x^2 + y^2)}{2}} \right ]^* \sqrt{\frac{1}{\pi}} \frac{i}{i + \delta_z}\exp{-i\frac{((x+\delta_x)^2 + y^2)}{2(i+\delta_z)} - i k z_R \delta_z} \nonumber \\
     &= \sum_\zeta^j  \sum_\xi^l c_\zeta^*  d_\xi^* \frac{1}{\pi} \frac{i \exp{-ikz_R \delta_z} \exp{\frac{i \delta_x^2}{2i+\delta_z}}}{i+\delta_z} \int dx \ x^\zeta \exp{-\frac{(x - \mu_x)^2}{2\sigma_{z_v}^2}} \int dy \ y^\xi \exp{-\frac{(y - \mu_y)^2}{2\sigma_{z_v}^2}},
\end{align}
with $\mu_x = i\delta_x / (2i + \delta_z), \mu_y = i\delta_y / (2i + \delta_z)$, and $\frac{1}{2\sigma_{z_v}^2} = \frac{1}{2} [ 1 + \frac{i}{i + \delta z_v}]$. We can use the formula for complex-valued Gaussian integrals,
\begin{align}
    \int dx \ \exp{-ax^2} = \sqrt{\frac{\pi}{a}}
\end{align}
since $\Re{a} \geq 0$ will always hold. We can now express each integral as the generating functions of the moments for a normal distribution and obtain the result
\begin{align}
   \braket{\phi_{j,l}}{\psi(x+\delta_x,y+\delta_y,\delta_z)} &= \frac{2i}{2i + \delta_z}  \exp{-ikz_R\delta_z} \exp{-\frac{i}{2(2i +\delta_z)}(\delta_x^2 + \delta_y^2)} \sum_\zeta^j c^*_\zeta \mathbb{E}[x^\zeta]\sum_\xi^l d^*_\xi \mathbb{E}[y^\xi],
\end{align}
with $\mathbb{E}[x_v^\zeta]$ (resp. $\mathbb{E}[y_v^\xi]$) the moments of the Gaussian distribution that can be computed recursively via
\begin{align}
    \label{def:recursive_moments}
    &\mathbb{E}[x_v^{\zeta+1}] = \mu_{x_v} \mathbb{E}[x_v^\zeta] + \zeta \sigma_{z_v}^2 \mathbb{E}[x_v^{\zeta-1}], \\
    &\mathbb{E}[x_v^0] = 1,  \quad \mathbb{E}[x_v^1] = \mu_{x_v} \nonumber.
\end{align}
Taking the absolute square of the inner product leads to the single source measurement probability
\begin{align}
\label{eq:three_sources_incoherent_ms_prob}
    \abs{\braket{\phi_{j,l}}{\psi(x + \delta x_v, y + \delta y_v, + \delta z_v)}}^2 = \frac{4}{4 + \delta z_v^2} \exp{-2 \frac{\delta x_v^2 + \delta y_v^2}{4+\delta z_v^2}} \mbox{$\abs{\sum_\zeta^j c^*_\zeta \mathbb{E}[x_v^\zeta]}^2$} \mbox{$\abs{\sum_\xi^l d^*_\xi \mathbb{E}[y_v^\xi]}^2$}.
\end{align}
We can further reduce deriving the derivatives w.r.t. the point source displacements $(\delta_x,\delta_y,\delta_z)$ Eq.~\eqref{eq:pdv_mode_dot_product}
\begin{align}
    \frac{\partial}{\partial_\theta} \abs{\braket{\phi_{j,l}}{\psi(x+\delta_x,y+\delta_y,\delta_z)}}^2 &= \left ( \frac{\partial}{\partial_\theta} \braket{\phi_{j,l}}{\psi(x+\delta_x,y+\delta_y,\delta_z)}^*\right ) \braket{\phi_{j,l}}{\psi(x+\delta_x,y+\delta_y,\delta_z)} \nonumber \\
    &+ \braket{\phi_{j,l}}{\psi(x+\delta_x,y+\delta_y,\delta_z)}^* \frac{\partial}{\partial_\theta} \braket{\phi_{j,l}}{\psi(x+\delta_x,y+\delta_y,\delta_z)}
\end{align}
to computing the derivative of the inner product. We first centralise the random variable $x \sim \cal{N} (\mu,\sigma^2)$ by $x=\mu + \sigma Z$, where $Z \sim \cal{N}$$(0,1)$ and re-express the recurrence relation Eq.~\eqref{def:recursive_moments} as a sum
\begin{align}
    \mathbb{E}[x^\zeta] = \mathbb{E}[(\mu + \sigma Z)^\zeta] = \sum_\alpha^\zeta \binom{\zeta}{\alpha} \mu^\alpha \sigma^{\zeta-\alpha} \mathbb{E}[Z^{\zeta-\alpha}],
\end{align}
where
\begin{align}
    \mathbb{E}[Z^{\zeta-\alpha}] = \begin{cases}
        \frac{(\zeta-\alpha)!}{2^{\frac{\zeta-\alpha}{2}} (\frac{\zeta-\alpha}{2})!} & \zeta-\alpha \text{ is even} \\
        0 & \text{else}
    \end{cases}.
\end{align}
The individual derivatives of the measurement mode overlap have the closed form expressions
\begin{align}
    \frac{\partial}{\partial \delta_x} \braket{\phi_{j,l}}{\psi(x+\delta_x,y+\delta_y,\delta_z)} &= \frac{2i}{2i + \delta_z}  \exp{-ikz_R\delta_z} \exp{-\frac{i \delta_y^2}{2(2i +\delta_z)}} \sum_\xi^l d^*_\xi \mathbb{E}[y^\xi]\sum_\zeta c^*_\zeta  \frac{\partial}{\partial \delta_x} \left [ \exp{-\frac{i\delta_x^2}{2(2i +\delta_z)}}\mathbb{E}[x^\zeta] \right ] \nonumber \\
    &=  \frac{2}{(2i + \delta_z)^2}  \exp{-ikz_R\delta_z} \exp{-\frac{i (\delta_x^2 + \delta_y^2)}{2(2i +\delta_z)}} \left ( \sum_\xi^l d^*_\xi \mathbb{E}[y^\xi] \right ) \left [ \delta_x \sum_\zeta c^*_\zeta \mathbb{E}[x^\zeta] \right. \nonumber \\
    & \left. + \sum_\zeta c^*_\zeta \sum^\zeta_\alpha \binom{\zeta}{\alpha} \alpha \mu_x^{\alpha-1} \sigma^{\frac{\zeta-\alpha}{2}}  \mathbb{E}[Z^{\zeta-\alpha}] \right ].
\end{align}
Similarly, by symmetry, we get the expression for the derivative in the $y$-dimension
\begin{align}
    \frac{\partial}{\partial \delta_y} \braket{\phi_{j,l}}{\psi(x+\delta_x,y+\delta_y,\delta_z)} &=  \frac{2}{(2i + \delta_z)^2}  \exp{-ikz_R\delta_z} \exp{-\frac{i (\delta_x^2 + \delta_y^2)}{2(2i +\delta_z)}} \left ( \sum_\zeta^j c^*_\zeta \mathbb{E}[x^\zeta] \right ) \left [ \delta_y \sum_\xi d^*_\xi \mathbb{E}[y^\xi] \right. \nonumber \\
    & \left. + \sum_\xi d^*_\xi \sum^\xi_\beta \binom{\xi}{\beta} \beta \mu_y^{\beta-1} \sigma^{\frac{\xi-\beta}{2}}  \mathbb{E}[Z^{\xi-\beta}] \right ].
\end{align}
Lastly, we have
\begin{align}
    \frac{\partial}{\partial \delta_z} &\braket{\phi_{j,l}}{\psi(x+\delta_x,y+\delta_y,\delta_z)} = \nonumber \\
    &\sum_\zeta^j c^*_\zeta \mathbb{E}[x^\zeta]\sum_\xi^l d^*_\xi \mathbb{E}[y^\xi] \frac{\partial}{\partial \delta_z} \left [ \frac{2i}{2i + \delta_z}  \exp{-ikz_R\delta_z} \exp{-\frac{i (\delta_x^2 + \delta_y^2)}{2(2i +\delta_z)}}  \right] \nonumber \\
    &+ \frac{2i}{2i + \delta_z}  \exp{-ikz_R\delta_z} \exp{-\frac{i (\delta_x^2 + \delta_y^2)}{2(2i +\delta_z)}} \frac{\partial}{\partial \delta_z} \left[ \sum_\zeta^j c^*_\zeta \mathbb{E}[x^\zeta]\sum_\xi^l d^*_\xi \mathbb{E}[y^\xi] \right] \nonumber \\
    &= \frac{4 - \delta_x^2 - \delta_y^2 -2i\delta_z + 2k z_R(2i+\delta_z)^2}{(2i + \delta_z)^3}\exp{-i k z_R \delta_z} \exp{-\frac{\delta_x^2 + \delta_y^2}{2(2i + \delta_z)}} \sum_\zeta^j c^*_\zeta \mathbb{E}[x^\zeta]\sum_\xi^l d^*_\xi \mathbb{E}[y^\xi] \nonumber \\
    &+ \frac{2i}{2i + \delta_z} \exp{-i k z_R \delta_z} \exp{-\frac{\delta_x^2 + \delta_y^2}{2(2i + \delta_z)}} \nonumber \\
    &\left [ \left ( \sum_\xi^l d_\xi^* \mathbb{E}[y^\xi] \right ) \sum_\zeta^j c^*_\zeta \sum_\alpha^\zeta \binom{\zeta}{\alpha} \frac{(\zeta - \alpha)!}{2^{\frac{\zeta-\alpha}{2}}(\frac{\zeta-\alpha}{2})!} \frac{i}{(2i+\delta_z)^2} \left ( \frac{\zeta - \alpha}{2} \mu_x^\alpha \sigma^{\frac{\zeta - \alpha}{2}-1} + \delta_x \alpha \mu_x^{\alpha-1}  \sigma^{\frac{\zeta - \alpha}{2}} \right ) \right ] \nonumber \\
    &+ \left [ \left ( \sum_\zeta^j c_\zeta^* \mathbb{E}[x^\zeta] \right ) \sum_\xi^l d^*_\xi \sum_\beta^\xi \binom{\xi}{\beta} \frac{(\xi - \beta)!}{2^{\frac{\xi-\beta}{2}}(\frac{\xi-\beta}{2})!} \frac{i}{(2i+\delta_z)^2} \left ( \frac{\xi - \beta}{2} \mu_y^\beta \sigma^{\frac{\xi - \beta}{2}-1} + \delta_y \beta \mu_y^{\beta-1}  \sigma^{\frac{\xi - \beta}{2}} \right ) \right ].
\end{align}

\begin{figure*}[h]
\centering
\subfloat[Fisher information for the estimation of the width parameter $\delta x$.]{\label{fig:incoh_FI_s}\includegraphics[width=0.45\columnwidth]{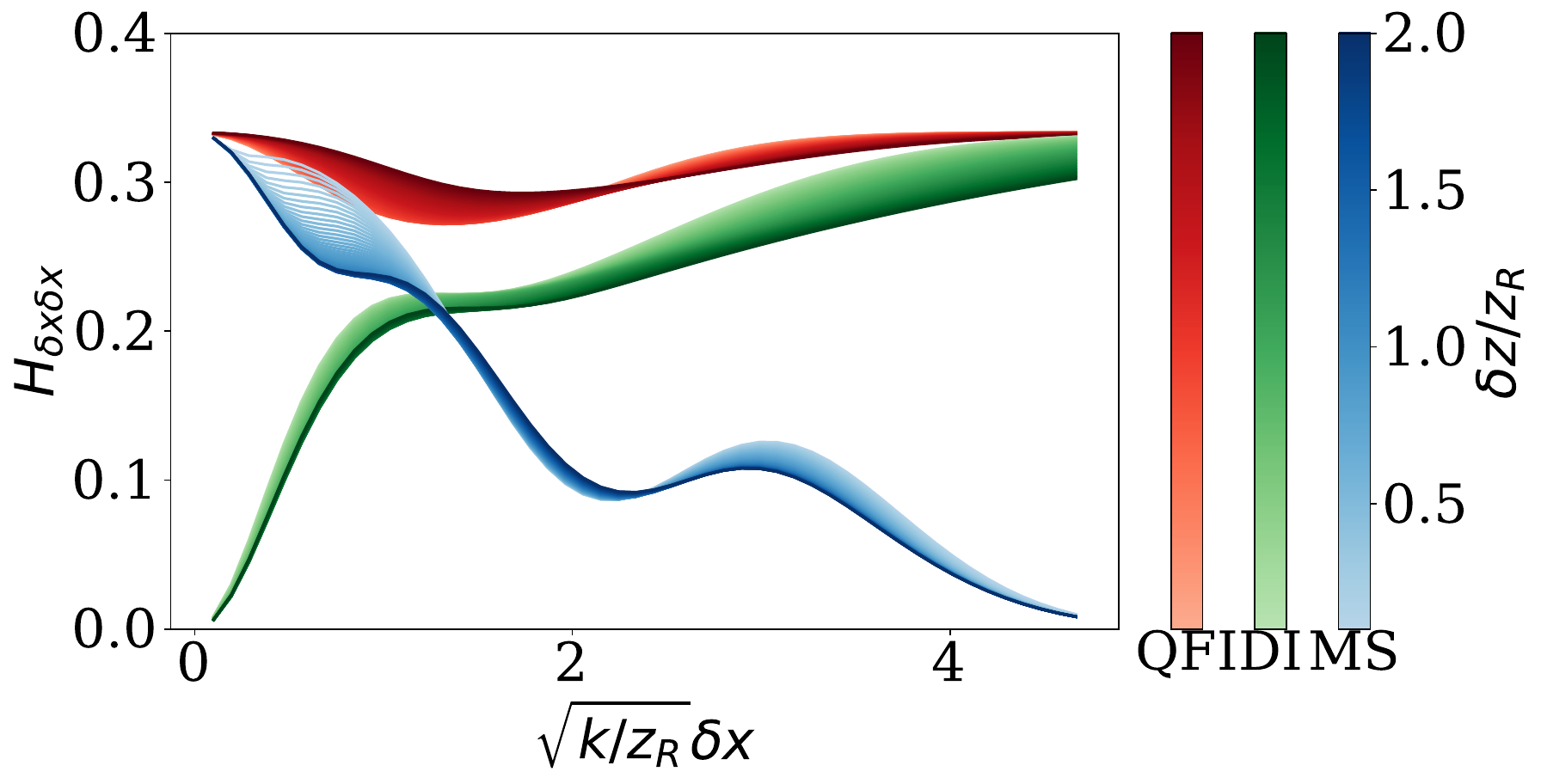}}\qquad
\subfloat[Individual mode contributions to the Fisher information for the estimation of the width parameter $\delta x$.]{\label{fig:incoh_MS_s}\includegraphics[width=0.45\columnwidth]{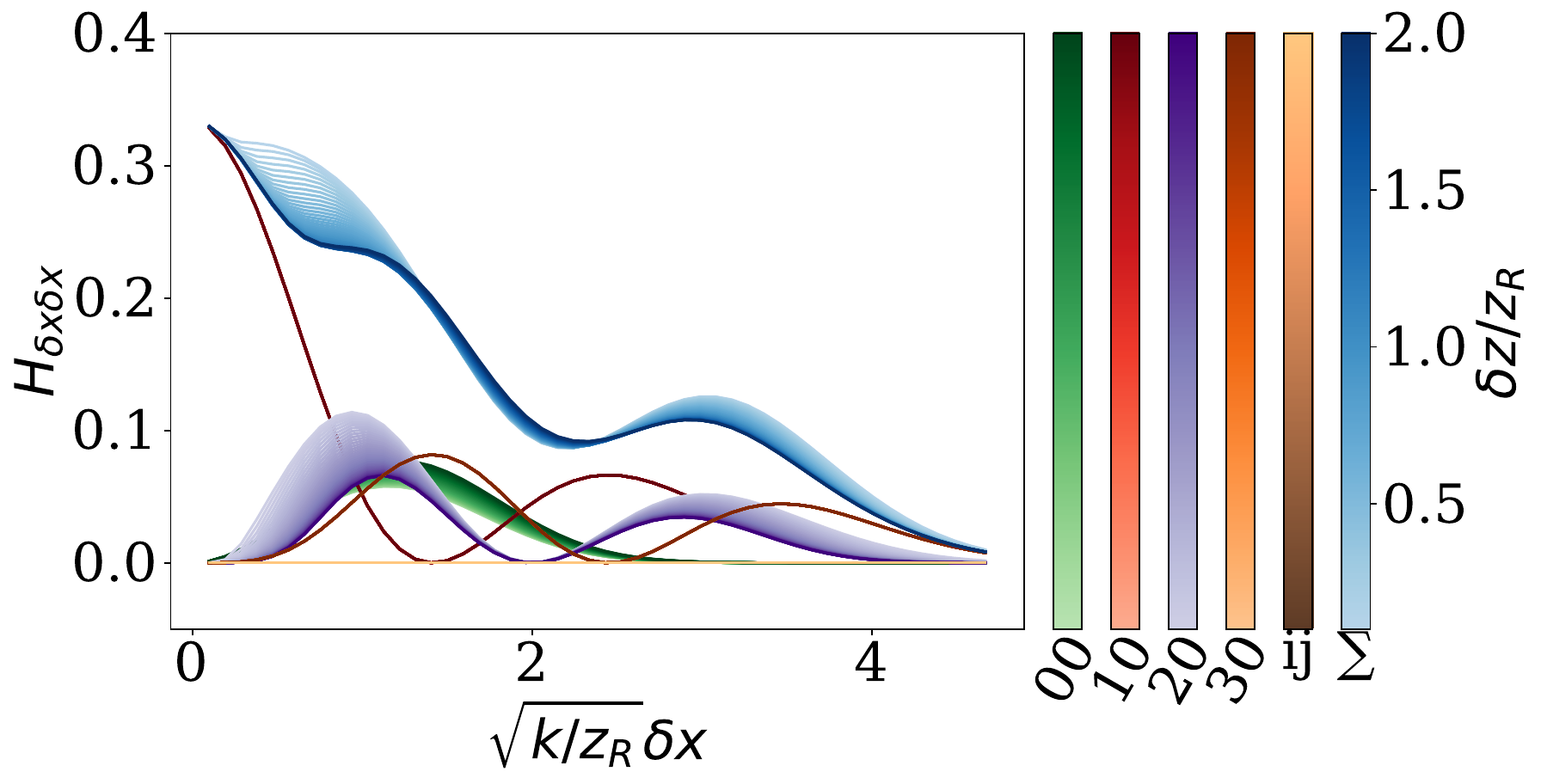}}\\
\subfloat[Fisher information for the estimation of the depth parameter $\delta z$.]{\label{fig:incoh_FI_p}\includegraphics[width=0.45\columnwidth]{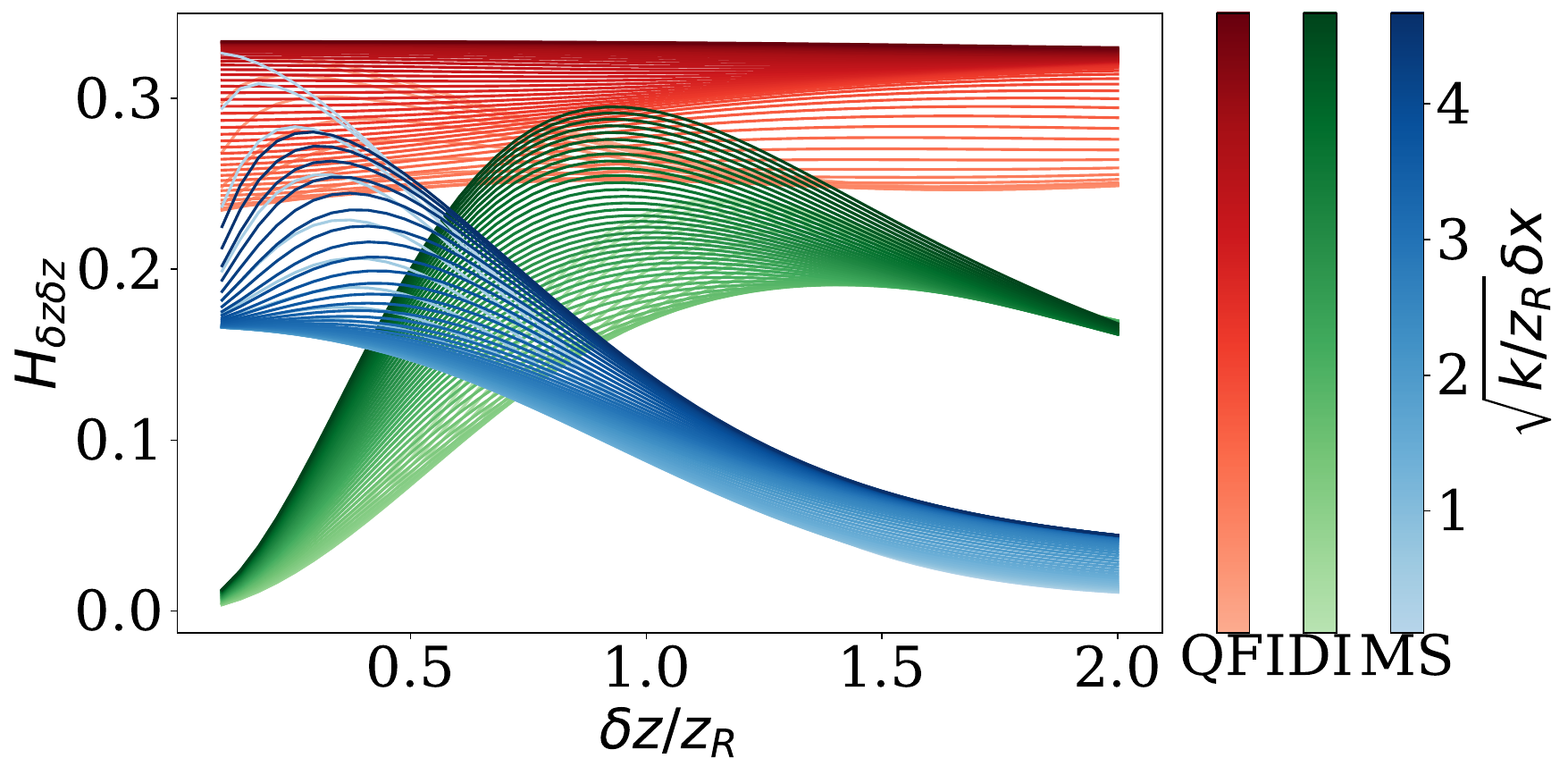}}\qquad
\subfloat[Individual mode contributions to the Fisher information for the estimation of the depth parameter $\delta z$.]{\label{fig:incoh_MS_p}\includegraphics[width=0.45\columnwidth]{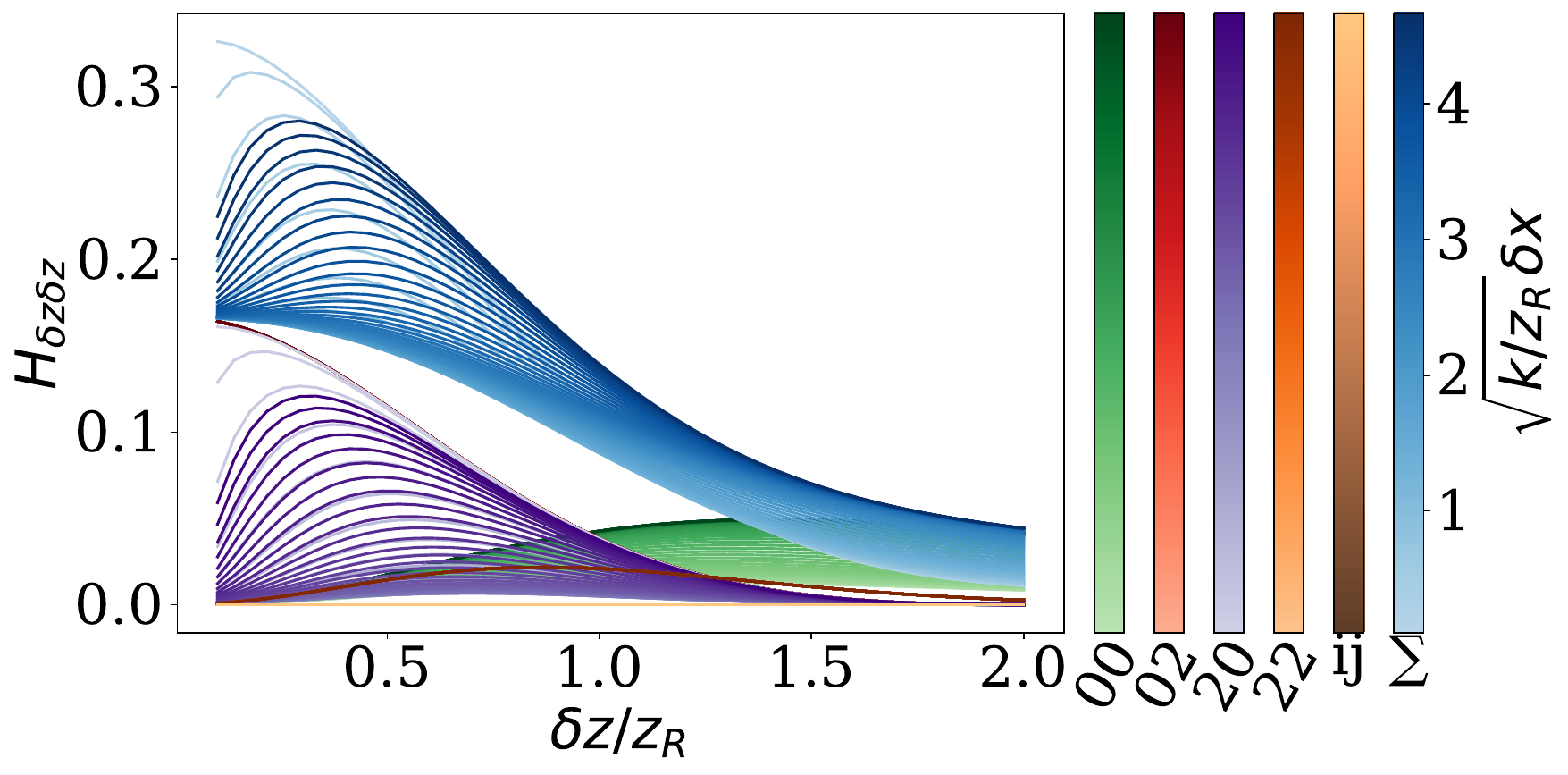}}%
\caption{Estimation of the parameters $\delta x$ [(a),(b)] and $\delta z$ [(c),(d)], for the incoherent three-source crack model depicted in Fig.~1 of the main text. Contributions of Hermite-Gaussian modes to the Fisher Information (Blue - MS), compared to the Quantum Fisher Information (Red - QFI) and the classical intensity measurement (Green - DI) [(a),(c)]. Individual Hermite-Gaussian modes with $(i,j)$ the 2D Hermite polynomial order, where for brevity we combine all the modes where the FIM matrix element is 0 (Brown - ij) and the last column is the sum of all the modes (Blue - $\sum$) [(b),(d)].}
\label{fig:supp4figs}
\end{figure*}

As an example, the Fisher information matrix elements for the estimation of the depth and width parameters of a surface crack modelled as in Fig.~1 of the main text are plotted in Fig.~\ref{fig:supp4figs}

\section{QCB Calculation} \label{app:QCB_calculation}

In this appendix, we describe the numerical methods used to calculate the quantum Chernoff bound and derive an analytical leading-order approximation for the fidelity between the density matrices.

\subsection{Numerical QCB Calculation} \label{app:QCB_calculation_numerical}

To numerically calculate the quantum Chernoff bound, we require an orthonormal basis to represent our density matrices $\rho_0$ and $\rho_1$. This is necessary to calculate the matrix powers $\rho_0^s$ involved in the formula. However, it is worth noting that the QCB is invariant under unitary conjugation, implying that any specific choice of basis should not impact our calculations. Since the point sources are continuous and non-orthogonal, we can use an orthonormal basis that describes the subspace that they span.

To find this subspace basis numerically, we utilise the Gram matrix $G = \braket{\psi_i}{\psi_j}$ of all four point sources $\ket{\psi_L}, \ket{\psi_R}, \ket{\psi_C}$ and $\ket{\psi_Z}$. As all the overlaps between the sources are contained within this matrix, it fully characterises their geometric relationships. By choosing a matrix $A$ such that $G=A^\dagger A$, we can use the columns of $A$ as vector representations of our point sources that have the same geometry. For simplicity, we choose $A = \sqrt{G}$, which exists since the Gram matrix is positive semidefinite. From the vectors we obtain, we can calculate finite dimensional numerical representations of our density matrices for given values of $\delta x$ and $\delta z$.

Having constructed numerical representations of $\rho_0$ and $\rho_1$, we can now calculate their eigenvalues $\lambda_i$, $\mu_j$ and eigenvectors $u_i$, $v_j$ \cite{audenaert2007discriminating}. The QCB is then calculated by minimising the trace
\begin{equation}
    \text{Tr}\left[\rho_0^s \rho_1^{1-s} \right] = \sum_{ij} \lambda_i^s \mu_j^{1-s} \abs{\braket{u_i}{v_j}}^2,
\end{equation}
where $s$ is minimised over the interval $[0,1]$:
\begin{equation}
\xi_Q = -\log \min_{0 \leq s \leq 1} \text{Tr} \left[ \rho_0^s \rho_1^{1-s} \right].
\end{equation}

\subsection{Analytical Fidelity Approximation} \label{app:fidelity_approximation}

For the leading-order expansion of the fidelity
\begin{equation}
F(\rho_0, \rho_1) = \text{Tr}\left[ \sqrt{\sqrt{\rho_0} \rho_1 \sqrt{\rho_0}}\right],
\end{equation}
we use a purification-based method from  \cite{lupo2020quantum}. A density matrix $\rho$ can be represented as ensemble of pure states
\begin{equation}
    \rho = \sum_j p_j \ketbra{\psi_j}{\psi_j}
\end{equation}
where each pure state $\ket{\psi_j}$ is defined using an orthonormal basis $\{\ket{i}\}$
\begin{equation}
\ket{\psi_j} = \sum_i \braket{i}{\psi_j} \ket{i}.
\end{equation}

The purification of $\rho$ is then written as
\begin{equation}
\ket{\Psi} = \sum_{ij} C_{ij} \ket{i} \ket{j},
\end{equation}
where $C_{ij} = \sqrt{p_j} \braket{i}{\psi_j}$ and $\{\ket{j}\}$ is an orthonormal basis representing each source.

For $\rho_0$ and $\rho_1$ we can calculate purifications $\Psi_0$ and $\Psi_1$ with corresponding matrices $C_0$ and $C_1$ respectively. The fidelity is then given by the formula
\begin{equation}
F(\rho_0, \rho_1) = \max_W |\bra{\Psi_0} I \otimes W \ket{\Psi_1}|.
\end{equation}
Using $C_0$ and $C_1$ this can then be reduced to a simple analytical expression
\begin{equation}
\begin{aligned}
F(\rho_0, \rho_1) =& \max_W \left| \sum_{jk} W_{jk} \sum_{i} (C_0)^*_{ij} (C_1)_{ik} \right| \\
=& \max_W \left| \text{Tr}[W^\top M] \right| = \|M\|_1,
\end{aligned}
\end{equation}
where $\|M\|_1 = \text{Tr}[\sqrt{M^\dagger M}]$ is the trace norm and $M = (C_0)^\dagger (C_1) = \sum_i (C_0)^*_{ij} (C_1)_{ik}$.
Through algebraic manipulation, we can express $M$ directly in terms of the overlaps of the ensembles $\rho_0 = \sum_j p_j \ketbra{\psi_j}{\psi_j}$ and $\rho_1  = \sum_k q_k \ketbra{\phi_k}{\phi_k}$
\begin{equation}
\begin{aligned}
M_{jk} =& \sum_i (C_0^*)_{ij} (C_1)_{ik}\\
= & \sum_{i} (\sqrt{p_j}\braket{i}{\psi_j})^*(\sqrt{q_k} \braket{i}{\phi_k})\\
= & \sqrt{p_j q_k} \sum_i \braket{\psi_j}{i} \braket{i}{\phi_k} = \sqrt{p_j q_k} \braket{\psi_j}{\phi_k}.
\end{aligned}
\end{equation}

By demonstrating that $M$ is a weighted overlap matrix, we show that it is also independent of the basis $\{\ket{i}\}$ used to represent each of the pure states $\ket{\psi_j}$ and $\ket{\phi_j}$.
Furthermore, the trace norm $\|M\|_1$ is also independent of the pure state ensemble used to represent the density matrix. This is due to changes in ensembles acting as right-multiplication by a unitary on the individual purification matrix $C' = CU$. Consequently, given a change of ensemble in $\rho_0$ and $\rho_1$ corresponds to $C'_0 = C_0U_0$ and $C'_1 = C_1U_1$, the resulting matrix is then described as  $M' = C_0'^\dagger C'_1 = U_0^\dagger C_0^\dagger C_1 U_1 = U_0^\dagger MU_1$. Since the trace norm is invariant under multiplication from unitaries on either side $\|M'\|_1 = \|U_0^\dagger M U_1 \|_1 = \|M\|_1$, the value of the fidelity is also invariant.

Using these properties, we can create a matrix $M$ using a convenient choice of basis. To do so we first define the following notation for the overlaps of our pure states
\begin{equation}
\begin{aligned}
\alpha =& \braket{\psi_L}{\psi_C} = \braket{\psi_R}{\psi_C} = \exp \left(-\frac{\delta x^2}{16} \right) \\
\beta =& \braket{\psi_L}{\psi_Z} = \braket{\psi_R}{\psi_Z} = \frac{2i}{\delta z + 2i} \exp \left(-\frac{i \delta x^2}{8(\delta z + 2i)} - i k z_R \delta z \right) \\
\gamma =& \braket{\psi_C}{\psi_Z} = \frac{2i}{\delta z + 2i} \exp \left( -i k z_R \delta z \right)
\end{aligned}
\end{equation}
and additionally,
\begin{equation}
\braket{\psi_L}{\psi_R} = \exp \left(-\frac{\delta x^2}{4} \right) = \alpha^4.
\end{equation}
To simplify the matrix $M$, we can recognise that both $\rho_0$ and $\rho_1$ contain the same states $\ket{\psi_L}$ and $\ket{\psi_R}$. Furthermore, since both states have the same overlap with $\ket{\psi_C}$ and $\ket{\psi_Z}$ due to symmetry, we introduce the two parity states
\begin{equation}
\begin{aligned}
\ket{+} =&\frac{\ket{\psi_L} + \ket{\psi_R}}{\sqrt{2 + 2\alpha^4}} \\
\ket{-} =& \frac{\ket{\psi_L} - \ket{\psi_R}}{\sqrt{2 - 2\alpha^4}}.
\end{aligned}
\end{equation}
These provide us with an alternative description of our quantum states
\begin{equation}
\begin{aligned}
\rho_0 =& \frac{(1+\alpha^4)\ketbra{+}{+} + (1-\alpha^4)\ketbra{-}{-} + \ketbra{\psi_C}{\psi_C}}{3} \\
\rho_1 =& \frac{(1+\alpha^4)\ketbra{+}{+} + (1-\alpha^4)\ketbra{-}{-} + \ketbra{\psi_Z}{\psi_Z}}{3}
\end{aligned}
\end{equation}
and by calculating $M$ from these ensembles, we obtain a helpful block-diagonal form.
\begin{equation}
M = \begin{bmatrix}
\frac{1+\alpha^4}{3} & 0 & \frac{\sqrt{2} \alpha}{3} \\
0 & \frac{1 - \alpha^4}{3} & 0 \\
\frac{\sqrt{2} \beta}{3} & 0 & \frac{\gamma}{3}
\end{bmatrix}
= \underbrace{\left[ \frac{1-\alpha^4}{3} \right]}_{M_1} \bigoplus \frac{1}{3} \underbrace{\begin{bmatrix} 1+\alpha^4 & \sqrt{2}\beta \\ \sqrt{2}\alpha & \gamma \end{bmatrix}}_{M_2}.
\end{equation}

This block-diagonal structure allows us to decompose the calculation into $\|M\|_1 = \|M_1\|_1 + \|M_2\|_1$ which simplifies our analysis.
Since $M_1$ is a scalar, $\|M_1\|_1$ reduces to its absolute value $|\frac{1-\alpha^4}{3}| = \frac{1-\alpha^4}{3}$ as $\alpha^4 \leq 1$. Substituting the leading-order approximation $\alpha^4 \approx 1 - \frac{\delta x^2}{4}$, we get $\|M_1\|_1 \approx \frac{\delta x^2}{12}$.
To calculate $\|M_2\|_1$, we can use the $2 \times 2$ matrix formula for the trace norm
\begin{equation}
\|A\|_1 = \sqrt{\text{Tr}\left[A^\dagger A \right] + 2 \abs{\operatorname{det}(A)}},
\end{equation}
along with leading-order approximations of the other variables $\beta$ and $\gamma$, to find that $\|M_2\|_1 \approx 1 - \frac{\delta x^2}{12} - \frac{\delta z^2}{24}$.

By combining these contributions, we obtain a leading-order approximation for the fidelity
\begin{equation}
\begin{aligned}
F(\rho_0,\rho_1) = \|M\|_1 \approx \frac{\delta x^2}{12} + 1 - \frac{\delta x^2}{12} - \frac{\delta z^2}{24} = 1 - \frac{\delta z^2}{24}.
\end{aligned}
\end{equation}

Finally, we can combine this result with Eq.~\eqref{eq:fidelity_bound}, using the small log approximation, to obtain an upper limit on the quantum Chernoff bound as reported in the main text,
\begin{equation}
    \xi_Q \leq \frac{\delta z^2}{12}.
\end{equation}

\end{document}